\magnification=\magstep1
\hfuzz=6pt
\baselineskip=16pt

$ $

\vskip 1in

\centerline{\bf Computational capacity of the universe}

\bigskip

\centerline{Seth Lloyd}

\centerline{d'Arbeloff Laboratory for Information Systems and Technology}

\centerline{MIT Department of Mechanical Engineering}

\centerline{MIT 3-160, Cambridge, Mass. 02139}

\centerline{slloyd@mit.edu}

\bigskip

{\bf 
Merely by existing, all physical systems register information.
And by evolving dynamically in time, they transform and process
that information.  The laws of physics determine the amount of 
information that a physical system can register (number of
bits) and the number of elementary logic operations that 
a system can perform (number of ops).  The universe is a physical
system.  This paper quantifies the amount of information that the
universe can register and the number of elementary operations that
it can have performed over its history.   The universe can have 
performed no more than $10^{120}$ ops on $10^{90}$ bits.
}

`Information is physical'$^1$.  This statement of Landauer has two
complementary interpretations.  First, information is registered and
processed by physical systems.  Second, all physical systems register
and process information.  The description of physical systems in terms
of information and information processing is complementary to the conventional
description of physical system in terms of the laws of physics.
A recent paper by the author$^2$ put bounds 
on the amount of information processing that can be performed by
physical systems.  
	
The first limit is on speed.  The Margolus/Levitin theorem$^3$
implies that the total number of elementary operations that a 
system can perform per second is limited by its energy:
${\rm \# ops/sec} \leq 2 E/\pi\hbar$, where $E$ is the system's
average energy above the ground state and $\hbar = 1.0545 \times 10^{-34}$ 
joule-sec is Planck's reduced constant.  As the presence of Planck's constant
suggests, this speed limit is fundamentally quantum mechanical,$^{2-4}$ and 
as shown in (2), is actually attained by quantum computers,$^{5-27}$     
including existing devices.$^{15,16, 21, 23-27}$  

The second limit is on memory space.  The total number of bits available 
for a system to process is limited by its entropy:
${\rm \# bits} \leq S/k_B \ln 2$, where $S$ is the system's
thermodynamic entropy and $k_B = 1.38 \times 10^{-23}$ joule/K
is Boltzmann's constant.$^2$  This limit, too, is attained by existing quantum
computers, which store one bit of information per nuclear spin or
photon polarization state.

The speed at which information can be moved from place to place
is limited by the speed of light, $c= 2.98 \times 10^8$ meters/sec.
This limit can be combined with the first two to limit the input/output
rate of computational systems.$^2$  The maximum rate at which information
can be moved in and out of a system with size $R$ is 
$\approx cS/k_B R$ (attained by taking all the information $S/k_B\ln2$ in the
system and moving it outward at the speed of light).  The ratio between
the maximum rate of information processing (ops/sec) and the
input/output rate (bits in/out per second) is $\approx k_B E R/ \hbar c S$,
a quantity shown by Bekenstein to be $\geq 1/2\pi$, with equality attained
for black holes.$^{28-30}$

Here, these three limits are applied to calculate the information processing
capacity of the universe.  In particular, I calculate both the total number
of bits available in the universe for computation, and the the number 
of elementary logical operations that can have been performed on those bits
since the universe began.  Perhaps not surprisingly, the total number of
bits ($ \approx 10^{90}$ in matter, $\approx 10^{120}$ if gravitation
is taken into account) and ops ($ \approx 10^{120}$) turn out to 
be simple polynomials in 
the fundamental constants of nature, Planck's constant $\hbar$, the speed 
of light $c$ and the gravitational constant $G=6.673 \times 10^{-11}$ 
meters$^3$/kilogram sec$^2$, together with the current age of the
universe.  These numbers of ops and bits can be interpreted in three
distinct ways:
\bigskip

\item{1.} They give upper bounds to the amount of computation
that can have been performed by all the matter in the universe since the
universe began.

\item{2.} They give lower bounds to the number of ops and
bits required to simulate the entire universe on a quantum computer.

\item{3.} If one chooses to regard the universe as performing
a computation, these numbers give the numbers of ops and bits in that
computation.

\bigskip\noindent  
The first two interpretations are  straightforward: 
even if all the matter in the universe were organized in
a computer, that computer could not register more information or
perform more elementary logical operations than calculated here.
Similarly, a quantum computer that accurately simulates the time evolution
of the universe requires at least this number of bits and must
perform at least this number of operations.$^{18-20}$ The third interpretation 
is potentially more controversial.$^{31}$  It is well established that
all degrees of freedom can be taken to register information and all 
physical dynamics transforms that information:
every nuclear spin can be thought of as registering a bit, and when 
the spin flips from clockwise to counterclockwise, the bit flips.  
Similarly, it is known that fundamental interactions between
elementary particles allow the performance of quantum logic
operations.$^{12-16}$
According to the third interpretation, every particle in the universe 
registers one or more bits, and every time those particles interact,
they perform one or more elementary operations.    
These interpretations will be discussed in greater detail below. 

Now let us calculate the computational capacity of the universe.  For
clarity of exposition, the calculation will not explicitly keep track of
factors of $1/2, \pi$, etc., that only affect the final results by an
order of magnitude or so (i.e., a factor of $2\pi$ will be ignored;
a factor of $(2\pi)^6$ will not be).  The expression $\approx$ will be used 
to indicate equality to within such factors.  In other words, 
$X\approx Y$ is equivalent to $\log X = \log Y + O(1)$.

To calculate the number of bits available and the number of ops that can
have been performed over the history of the universe requires a model
of that history.  The standard big-bang model will be used here.$^{32}$ In
this model the universe began $\approx 10^{10}$ years ago in
what resembled a large explosion (the big bang).  Since the big bang,
the universe has expanded to its current size.  The big bang model
successfully explains a number of features of the current universe,
such as its ongoing expansion and the existence of the microwave background.
Discussions of the amount of computation that can have been performed in the 
pre-big bang universe are necessarily speculative: the well-established 
inflationary scenario will be used to investigate computation in
this regime.$^{33-34}$  

For most of its history, the universe has been
matter-dominated -- most of the energy is in the form of matter.
As will be seen below, most of the computation that can have taken
place in the universe occurred during the matter-dominated phase.
Accordingly, begin with the computational capacity of the
matter-dominated universe, and then work back through the 
radiation-dominated and inflationary universe.

\bigskip\noindent {\bf 1. Number of ops}

First, investigate the number of elementary logic operations that can
have been performed.  As noted above, the maximum number of operations
per second that can be performed by a physical system is proportional
to its energy.  This result follows from the Margolus-Levitin theorem, 
which states that the minimum time required for a physical system to
move from one state to an orthogonal state is given by 
$\Delta t = \pi \hbar/2 E$, where $E$ is the average energy of
the system above its ground state.$^3$ 

The Margolus-Levitin theorem is a mathematical expression of the
familiar fact that energy must be invested to change a system
such as a quantum bit from one state to another.
As an example, consider the process of flipping a bit in an NMR
quantum computer.  In NMR quantum computing, each nuclear spin in
a molecule registers a quantum
bit, with spin up identified with $0$ and spin down identified
with $1$.  To flip a quantum bit, apply a magnetic field.   
The spin precesses about the magnetic field: when the spin has precessed
by an angle $\pi$, the state was spin up has become spin down and
{\it vice versa}.  In terms of bits, $0$ has become $1$ and
$1$ has become $0$: flipping the spin flips the bit.
It is straightforward to verify that the spin/bit flipping process
takes place in time $\Delta t = \pi \hbar /2 E = \pi/\omega_L$, where 
$E$ is the average energy of the spin's interaction with the magnetic
field and $\omega_L$ is the spin's Larmor frequency.  Two-qubit quantum
logic operations obey the same bound, with $E$ the average energy of
interaction between the qubits.   Note that while energy must be
invested in the spin-field interaction to flip the bit, it need
not be dissipated.

The Margolus-Levitin bound also holds for performing many logic
operations in parallel.  If energy $E$ is divided up among $N$
quantum logic gates, each gate operates $N$ times more slowly
than a single logic gate operating with energy $E$, but the maximum
total number of operations per second remains the same.

Now apply these results to the universe as a whole.
In the matter-dominated universe, the energy within a
co-moving volume is approximately
equal to the energy of the matter within that volume
and remains approximately constant over time.  (A co-moving volume
is one that is at rest with respect to the microwave background, and
that expands as the universe expands.)  Since the energy remains constant,
the number of ops per second that can be performed by the matter in a
co-moving volume remains constant as well.  The total volume of the universe
within the particle horizon is $\approx c^3 t^3$, where $t$ is the age
of the universe.  The particle horizon is the boundary between the part
of the universe about which we could have obtained information over the
course of the history of the universe and the
part about which we could not.  In fact, the horizon is currently somewhat 
further than $ct$ away, due to the ongoing expansion of the universe,
but in keeping with the approximation convention adopted above we will
ignore this factor along with additional geometric factors in estimating
the current volume of the universe.

The total number of ops per second that can be performed in the
matter-dominated universe is therefore
$ \approx \rho c^2 \times c^3 t^3/\hbar$, where
$\rho$ is the density of matter and $\rho c^2$ is the energy
density per unit volume.  Since the number of ops per second in a
co-moving volume is constant, and since the universe has been matter
dominated for most of its history, we have
$$ {\rm \# ops} \approx \rho c^5 t^4/\hbar. \eqno(1)$$
Insertion of current estimates for the density of the universe
$\rho \approx 10^{-27} {\rm kg/m^3} $ and the age of the universe
$t \approx  10^{10}$ years, we see that the universe could have
performed $\approx 10^{120}$ ops in the course of its history.

A more revealing form for the number of ops can be obtained by noting 
that our universe is close to its critical density.  If the
density of the universe is greater than the critical density, it
will expand to a maximum size and then contract.  If the density
is less than or equal to the critical density, it will expand forever.
The critical density can be estimated by a simple, non-relativistic
argument$^{32}$. 
At the critical density, the kinetic energy of a galaxy at distance
$R$ is equal to its gravitational energy, so that the galaxy has
just enough energy to move to infinite distance.
Galaxies at distance $R$ are moving away from
our galaxy at a rate $\approx H R$, where $H \approx 1/t $ is the Hubble
constant. ($H$ is somewhat smaller than $1/t$ and is not in
fact a constant, since the expansion is slowing over time.  But $H \approx 
1/t$ to within the approximation convention adopted here.)  
The kinetic energy of a galaxy of mass $m$ at distance $R$ is therefore
equal to $mH^2R^2/2$, and its potential energy is $4\pi G m\rho R^2/3$.
Equating the galaxy's kinetic energy with its potential energy yields
the critical density $\rho_c = 3H^2/8\pi G \approx 1/Gt^2$.

So for a matter-dominated universe at its critical density, the total
number of ops that can have been performed within the horizon at time
$t$ is
$$ {\rm \# ops} \approx \rho_c c^5 t^4/\hbar \approx t^2 c^5/G\hbar
= (t/t_P)^2 . \eqno(2)$$
Here $t_P = \sqrt{G\hbar/c^5} = 5.391\times  10^{-44} $ seconds is the Planck 
time --- the time scale at which gravitational effects are of the same order as
quantum effects.  

In retrospect, it should come as no surprise that the number of 
elementary operations that can have been performed within the horizon
over the lifetime of a universe at critical density should be some
polynomial in $t/t_P$.  The number of operations per second is a function
of the critical density $\rho_c$ and the age of the universe $t$, together
with the fundamental constants.  The total number of operations is a 
dimensionless number, and so should be a function of $t/\tau$, where
$\tau$ is a timescale that is a function of $\hbar, c$, and $G$. 
But the only such timescale is the Planck time.  The reason that the
number of ops performed within the horizon grows as $(t/t_P)^2$ rather
than simply $t/t_P$ is that the total amount of matter within the
horizon is growing with time: as the universe gets older, more and
more galaxies come into view.

The previous paragraphs calculate the maximum number of operations that could
be performed by the matter in the universe.  What about the number of
elementary operations that could be performed by the gravitational
field?  Care must be taken in estimating this number, as the Margolus-Levitin
theorem that allows the calculation of the number of operations performed
by a physical system is strictly quantum-mechanical in nature, and
no complete quantum theory of gravity currently exists.  With this
caveat in mind, however, one can still provisionally apply the theorem, which as
noted above states that the number of operations per second that 
can be performed by a physical system is bounded above by
$\approx E/\hbar$, where $E$ is the system's average energy above
the ground state energy.  The gravitational energy in the universe
is negative and equal in magnitude to the positive energy of the
matter in the universe, so that the total energy of the universe
is zero.  Since the Margolus-Levitin theorem enjoins the
calculation of the number of ops per second using the average energy
above the lowest energy that appears in the system's quantum state, 
the number of elementary
operations that can be performed using gravitational energy is exactly
equal to the number that can be performed using the energy in matter.
Including the number of operations that can have been performed by 
gravitational energy (however one might accomplish such a feat!) then
changes the overall number of operations by a factor of two.  Even
including gravitational energy, the total number of operations that
can have been performed in the universe is $\approx (t/t_P)^2
\approx 10^{120}$.

It is instructive to compare the total number of operations that
could have been performed using all the matter in the universe with
the number of operations that have been performed by conventional computers.
The actual number of elementary operations performed by all man-made
computers is of course much less than this number.  Because of
Moore's law, about half of these elementary operations have been
performed in the last two years.  Let us estimate the total
number of operations performed by man-made computers, erring
on the high side.  With $\approx 10^9$ computers
operating at a clock rate of $\approx 10^9$ Hertz performing
$\approx 10^5$ elementary logical operations per clock cycle over
the course of $\approx 10^{8}$ seconds, all the man-made computers 
in the world have performed no more than
$\approx 10^{31}$ ops over the last two years, and no more than approximately
twice this amount in the history of computation.  

What is the universe computing?
In the current matter-dominated universe most of the known energy is locked
up in the mass of baryons.  If one chooses to regard the universe
as performing a computation, most of the elementary operations in
that computation consists of protons, neutrons (and their constituent 
quarks and gluons), electrons and photons moving from place to place 
and interacting with each other according to the basic laws of physics.
In other words, to the extent that most of the universe is performing
a computation, it is `computing' its own dynamical evolution.  Only a small 
fraction of the universe is performing conventional digital computations.

\bigskip\noindent{\bf 2. Number of bits}

Information is registered by physical systems, and all physical
systems can register information.$^1$  The amount of information,
measured in bits,
that can be registered by any physical system is equal to the 
logarithm to the base 2 of the number of distinct quantum states
available to the system given its overall energy, volume, electric
charge, etc.$^2$  In other words, $I = S/k_B \ln 2$, where $S$ is the
maximum entropy of the system and $k_B = 1.38 * 10^{-23}$ joule/$K$
is Boltzmann's constant.

To calculate the number of bits that can be registered by the universe
requires a calculation of its maximum entropy, a calculation familiar
in cosmology. The maximum entropy in the matter-dominated universe would be 
obtained by converting all the matter into radiation.  (Luckily for us, we
are not at maximum entropy yet!)  The energy per unit volume is
$\rho c^2$.  The conventional equation for black-body radiation
can then be used to estimate the temperature $T$ that would be
obtained if that matter were converted to radiation at temperature $T$:
$ \rho c^2 = (\pi^2 /  30 \hbar^3 c^3) (k_B T)^4 \sum_\ell n_\ell. $
Here $\ell$ labels the species of effectively massless particles
at temperature $T$ (i.e., $m_\ell c^2 << k_B T$), and $n_\ell$
counts the number of effective degrees of freedom per species:
$n_\ell =$ (number of polarizations) $*$ (number of 
particles/antiparticles) $*$ 1 (for bosons) or 7/8 (for fermions).
Solving for the temperature for the maximum entropy state gives
$k_B T = (30 \hbar^3 c^5 \rho / \pi^2 \sum_\ell n_\ell)^{1/4}$.
The maximum entropy per unit volume is $S/V = 4\rho c^2/3T$.
The entropy within a volume $V$ is then 
$ S = (4k_B/3) ( \pi^2 \sum_\ell n_\ell /30)^{1/4} (\rho c/\hbar )^{3/4}
V^{1/4} $.  The entropy depends only weakly on the number of
effectively massless particles.  

Using the formula $I = S/k_B \ln 2$
and substituting $\approx c^3 t^3$ for the volume of the universe
gives the maximum number of bits available for computation:
$$ I \approx (\rho c^5 t^4/\hbar)^{3/4} = ( {\rm \# ops} )^{3/4} \eqno(3)$$   
The universe could currently register $\approx 10^{90}$ bits.
To register this amount of information requires every degree of freedom
of every particle in the universe. 

The above calculation estimated only the amount of information that
could be stored in matter and energy and did not take into account 
information that might be stored on gravitational degrees of freedom.
The calculation of the amount of information that can be stored
by a physical system is intrinsically quantum mechanical.  Accordingly,
just as in the calculation of the number of operations that could
be performed by the gravitational field, an accurate account of the
amount of information that can be registered gravitationally must await
a full quantum theory of gravity.  With this caveat in mind, note
that the Bekenstein bound$^{28-30}$ together with
the holographic principle$^{35-37}$ implies that the maximum amount of 
information that can be registered by any physical system, including 
gravitational ones, is equal to the area of the system divided by the the 
square of the Planck length, $\ell_P^2 = \hbar G/c^3$.  This limit is in fact
attained by black holes and other objects with event horizons.  
Applying the Bekenstein bound and the 
holographic principle to the universe as a whole implies
that the maximum number of bits that could be registered by the universe
using matter, energy, and gravity is $\approx c^2 t^2/\ell_P^2 = t^2/t_P^2$.
That is, the maximum number of bits using gravitational degrees of freedom
as well as conventional matter and energy is equal to the maximum number
of elementary operations that could be performed in the universe,
$\approx 10^{120}$.

Not surprisingly, existing man-made computers register far fewer
bits.  Over-estimating the actual number of bits registered in 2001, as
above for the number of ops, yields $\approx 10^9$ computers, each registering
at $\approx 10^{12}$ bits, for a total of $ \approx 10^{21}$ bits.

\bigskip\noindent{\bf 3. Large Numbers}

Three quarters of a century ago, Eddington
noted that two large numbers that characterize our universe happen
to be approximately equal.$^{38}$  In particular, the ratio between
the electromagnetic force by which a proton attracts an electron and
the gravitational force by which a proton attracts an electron
is $\alpha = e^2/Gm_em_p \approx 10^{40}$.  Similarly, the
ratio between the size of the universe and the classical size of
an electron is $\beta = ct/(e^2/m_ec^2) \approx 10^{40}$.
The fact that these two numbers are approximately equal is
currently regarded as a coincidence, although Dirac constructed
a theory in which they are equal by necessity.  Dirac's theory
is not favored by observation.

A third large number, the square root of the number of baryons
in the universe, $\gamma = \sqrt{\rho c^3 t^3/m_p}$ is also
$\approx 10^{40}$.  This is not a coincidence given the values
of $\alpha$ and $\beta$: $\alpha \beta \approx \gamma^2$
in a universe near its critical density
$\rho_c \approx 1/Gt^2$. 

The astute reader may have noted that the number of operations
that can have been performed by the universe is approximately equal
to the Eddington-Dirac large number cubed.  
In fact, as will now be shown, the number of ops is necessarily
approximately equal
to $\beta \gamma^2 \approx \alpha\beta^2 \approx 10^{120}$.
This relation holds true whether or not
$\alpha \approx \beta \approx \gamma$ is a coincidence. 
In particular, 
$$\beta\gamma^2 = (\rho c^5 t^4/\hbar) (\hbar c/e^2) (m_e/m_p)
= {\rm \# ops} * (137/1836)
\eqno(4a)$$
Similarly,
$$\alpha\beta^2 = (t/t^P)^2 (\hbar c/e^2) (m_e/m_p) 
= {\rm \# ops} * (137/1836)
\eqno(4b)$$
That is, the number of ops differs from the Eddington-Dirac large
number cubed by a factor of the fine structure constant times
the proton-electron mass ratio.  Since the number of ops is
$\approx 10^{120}$, as shown above, and the fine-structure constant
times the  proton-electron mass ratio is $\approx 10$, the
number of ops is a factor of ten larger than the Eddington-Dirac
large number cubed.  

In other words, whether or not the approximate equality embodied
by the Eddington-Dirac large number is a coincidence, the fact
that the number of operations that can have been 
performed by the universe is related to this large number is not.
These numerical relations are presented here for the sake of completeness.
The author encourages the reader not to follow in Dirac's footsteps
and take such numerology too seriously.

\bigskip
\noindent{\bf 4. Computation in the radiation-dominated universe}

The energy density of the universe became dominated by matter about
$700,000$ years after the big bang.  Before that time (with the exception
of the first fraction of a second), the energy density of the universe
was dominated by radiation, i.e., particles whose mass energy $mc^2$
is much less than their kinetic energy $k_B T$.  In the radiation-dominated
universe, the total energy in a co-moving volume is not constant.  As
the universe expands, the wavelength of the radiation is stretched by
the scale factor $a$, which increases as $t^{1/2}$, so that the energy in
an individual photon or particle of radiation is decreased by the
same factor.  Accordingly, if the energy in the radiation-dominated
universe available for computation at time $t_1$ is $E$, the energy available
at an earlier time $t_0$ is $E(t_1/t_0)^{1/2}$.  Note that this
result is independent of the total number of species of massless
particles as long as no phase transition takes place in which a species of
particle goes from being effectively massless to massive between
$t_0$ and $t_1$. 

In other words, as $t\rightarrow 0$, the total energy available for
computation in the radiation-dominated universe diverges.  Rephrased
in computational terms, the number of operations per second that can
be performed in the radiation-dominated universe diverges as $t\rightarrow 0$.
At first, this fact might seem to imply that the total number of operations
performed in the radiation-dominated universe is formally infinite.
However, as will now be seen, this is not the case.

The total number of operations that can have been performed within a
co-moving volume is 
$$ {\rm \# ops} = (2E/\pi\hbar) \int_{t_0}^{t_1} (t_1/t)^{1/2} dt
= (4E/\pi\hbar)(t_1 - \sqrt{t_1t_0}). \eqno(5)$$
But since $2(t_1-t_0) \geq 2(t_1 - \sqrt{t_1t_0}) \geq t_1 - t_0 $,
the total number of operations that can have been performed in the
radiation-dominated universe over a time interval $t_1 - t_2$ in 
a co-moving volume with energy $E$ at time $t_1$ is no more than
twice the number that can have been performed in the matter-dominated
universe in a co-moving volume with the same energy over the same
amount of time.  Accordingly, the formulae for the number of operations 
derived above for the matter-dominated universe also hold for the
radiation-dominated universe to within the accuracy of our approximation.

This result --- that the amount of information processing in the 
radiation-dominated universe is finite even as $t\rightarrow 0$ --- provides
a partial answer to a question posed by Dyson,$^{39}$ who addressed
the problem of the amount of information processing possible 
during the `big crunch.'
The big crunch is the time inverse of the big bang,
in which a closed universe contracts to a singularity.  The results
derived here for the number of operations and number of bits available
for computation hold equally for the big bang and the big crunch.
Equation (5) shows that the amount of computation that can be performed
in the radiation-dominated universe is finite, whether during the
big bang or the big crunch.  To perform an infinite number of operations 
would require that the amount of energy in a co-moving volume goes
to infinity at a rate $1/t$ or faster as $t\rightarrow 0$.  As will be seen
below, quantum gravitational considerations make it unlikely that an 
infinite number of ops can be performed as $t\rightarrow 0$.

Now calculate the number of bits available in the radiation-dominated
universe.  The calculation is almost identical to that leading to equation
(3), with the sole difference that here the dominant form of energy
(the radiation) is already assumed to be in a maximum entropy state,
unlike the matter in the matter-dominated universe.  The number of bits
available is 
$$I = S/k_B \ln 2 = 4E/3\ln 2 k_B T = (4/3\ln 2) ({\rm \# ops})^{3/4}
D^{1/4}, \eqno(6)$$
where $D=(\pi^2/30) \sum_\ell n_\ell$ (recall that $n_\ell$ is the
number of effective degrees of freedom for the $\ell$'th species
of effectively massless particle).  As long as $k_B T$ is below
the grand unification threshold of $2 \times 10^{16}$ GeV,
($T\approx 10^{29} K$), then $D^{1/4} \approx 1$, and the number of
bits available for computation is  $\approx {\rm \# ops}^{3/4}$,
just as in the matter-dominated universe.  If one includes the number
of bits that might be stored in gravitational degrees of freedom,
then the total number of bits available is $\approx (t/t_P)^2 
\approx {\rm \# ops}$, just as in the matter-dominated universe.

The formulae for the number of elementary operations and number of bits
available are essentially the same for the radiation-dominated universe 
and the matter-dominated universe.  That is not to say that the
radiation- and matter-dominated universes are computationally similar
in all respects.  The matter-dominated universe is far from thermal
equilibrium: most of the energy is in the form of free energy, which
is available for conversion to useful work via, e.g., nuclear fusion.
Most of the matter is in the form of the quarks and gluons that provide 
the structural stability of baryons.   In the radiation-dominated universe,
by contrast, most of the energy is in the form of effectively
massless particles at thermal equilibrium so that there 
is little free energy available and structural stability is hard to maintain.
In other words, the matter-dominated universe is a much more friendly 
environment for conventional digital computation, not to mention for
life as we know it.

\bigskip\noindent{\bf 5. Computation in the inflationary universe}

The results derived above put limits on the amount of computation 
that can have taken place in the universe from the Big Bang on.  The
dynamics of the Big Bang is governed by the Standard Model of elementary
particles, which is well-established to temperatures of at least
$k_B T \approx 100$ GeV, corresponding to a time of no more than $10^{-10}$
seconds after the universe began.  How far back the Big Bang extends,
and what happened before the Big Bang is not precisely known.  In
addition, although the Big Bang model accurately describes features
of the observed universe such as its expansion, the abundances of light
elements, and the cosmic microwave background, it fails to account for
the universe's observed flatness, homogeneity, and isotropy (as well as 
other features such as the scarcity of magnetic monopoles).  
Put more colloquially, the Big Bang model does not account for the
observed fact that the night sky looks qualitatively the same in
all directions: in the standard Big Bang model, sufficiently distant
pieces of the observed universe were never in causal connection and
so have no particular reason to be similar.

This apparent difficulty is resolved by recently developed 
inflationary cosmologies.$^{33-34}$  In these cosmologies, the scale factor
for the universe grows at an exponential rate for some period
of time, effectively `stretching out' the universe from an
initially small, causally connected volume to a large, causally
unconnected volume.  Since the observed universe now originates
from a causally connected volume, its homogeneity and isotropy
can be explained (as well as its flatness, the lack of magnetic
monopoles, etc.).  The details of the inflationary scenario depend on physics 
beyond the standard model, but its predictions are sufficiently compelling
that it has been adopted by many cosmologists as a promising
provisional theory of the very early universe.

The amount of computation that can have been performed during the inflationary
universe is readily calculated.  Inflation occurs when the scale
factor $a$ of the universe is accelerating: $\ddot a > 0$.
The Hubble parameter $H = \dot a/a$ measures the rate of expansion
of the universe, as above.  For simplicity, concentrate on a
universe at critical density.  For such a universe, 
$H = \sqrt{8\pi G \rho_c/3c^2}$, as above.  In the very early universe,
the energy density $\rho_c$ could be supplied by a scalar field
(the `inflaton') or by an effective cosmological constant.  During
inflation, the scale factor grows exponentially at a rate $H$.
Such a universe locally appears to be a De Sitter space and possesses
an horizon at distance $\approx c/H$: an inertial observer sees all matter
accelerating away at a rate proportional to its distance.  The energy
within the horizon is $\approx \rho_c c^3/H^3$ and the total number
of elementary logical operations that can be performed per second
within the horizon is $\approx \rho_c c^3/H^4 \hbar \approx  1/ t_P^2 H$.  
Similarly, the total
number of ops that can be performed within the horizon in a time
$t\approx 1/H$ is $ \approx 1/ t_P^2 H^2$.  In other words,
the amount of computation that can be performed within
the horizon in the Hubble time is $t^2/ t_P^2 $, as in the critical 
matter-dominated and radiation-dominated universes.
By the holographic principle, the maximum information that can be stored 
within the horizon is equal to the horizon area $\approx c^2/H^2$ divided 
by the Planck length squared.  So the number of bits that can be registered
within the horizon is no greater than $t^2/t_P^2 $, as well. 

As just shown, the formulae derived for number of bits and number of
ops in a universe at critical density hold during inflation.
The horizon scale is thought to go from the Planck scale at inflation
to a scale between $10^{10\pm 6}$ times larger at the end of inflation.
Accordingly, the universe can have performed no more than $10^{20\pm 12}$
ops on the same number of bits within the horizon during inflation (the
large uncertainties reflect the current theoretical uncertainties in
the inflationary scenario).  Inflation is followed by a period
of matter creation and reheating leading up to the start of the hot
Big Bang.  The formulae derived for the universe at critical density
hold for this stage as well.

Even though the amounts of computation that can have been performed
during inflation are relatively small, there is one highly significant
information-based process that occurs during inflation.  Because the
scale factor for the universe increases by $\approx 50$ $e$-foldings
($e^{50}$), the volume of the initially causally connected section
of the universe increases by $\approx 150$ $e$-foldings.  The resulting
volume consists of causally unconnected sectors (separated by horizons
as long as inflation persists).  Although these sectors cannot participate
in a joint computation (they cannot communicate with eachother), they
have a much higher free energy and so a much higher maximum entropy 
than the initial volume.  In other
words, the exponential expansion can be regarded as a massive 
`bit-creation' process. Indeed, Zizzi has suggested that inflation be 
regarded as adding new bits to a quantum computer, with the computation
provided by the action of quantum gravity.$^{35-36}$ Recent results suggest
that quantum gravitational systems can be thought of in terms of
quantum computation,$^{2, 35-37}$ a striking example of Wheeler's
slogan, `It from Bit.'$^{40}$ Quantum gravity provides a
maximum energy density given by the Planck scale and a maximum
bit density supplied by the holographic principle.  Accordingly,
the amounts of computation that can be performed in a quantum
gravitational context are finite.  Indeed, the fundamentally Planck-scale
character of the limits derived above suggests that they hold equally
in quantum gravity.  But whether or not these limits hold for
the first instant of the universe's existence
is a question, like many raised here,
whose answer must await the construction of a consistent theory of quantum 
gravity.  

\bigskip\noindent{\bf 6. Discussion}

The above sections calculated how many elementary logical
operations that can have been performed on how many bits during
various phases of the history of the universe.  As noted above,
there are three distinct interpretations of the numbers calculated.
The first interpretation simply states that the number of ops and
number of bits given here are upper bounds on the amount of computation
that can have been performed since the universe began.  This interpretation
should be uncontroversial: existing computers have clearly performed far
fewer ops on far fewer bits.  As Moore's law progresses and as computers use 
fewer degrees of freedom per bit and less energy per logic operation,  
the number of bits and ops available will increase.  Existing quantum
computers already use the minimum number of degrees of freedom per
bit and the minimum energy per operation.  The maximum amount of
computing power that will eventually be available to mankind is a
question of considerable technological and social interest.$^{41}$
Of course, that maximum computing power is likely to remain far below
the amounts calculated for the universe as a whole.

The second interpretation notes that the numbers calculated give a lower
bound on the number of bits and the number of operations that must be performed 
by a quantum computer that performs a direct simulation of the universe.
This interpretation should also be uncontroversial: quantum computers
have been shown to be efficient simulators of any physical system that
evolves according to local interactions.$^{18-20}$  
In particular, a quantum computer
can simulate the time evolution of any quantum system to an arbitrary
degree of accuracy using the same number
of degrees of freedom as the system, and a number of operations that is
proportional to the time over which the system evolves.  The universe is
a physical system, and to simulate the time evolution of the universe,
a quantum computer needs at least the same number of bits, 
and a number of operations
at least as great as the total angle $\int E dt/\hbar$ evolved by the 
system's state in Hilbert space.  Open questions include the efficiency
and robustness of such simulations.  For example, how much overhead in
terms of ops and bits is required to simulate the standard model of
elementary particles?  This is a question that could be answered using
current techniques.  How hard is it to simulate quantum gravity on a quantum
computer?  This is a question whose answer must await the construction
of a self-consistent theory of quantum gravity.  But the fact that quantum
computers can efficiently simulate any theory that is locally finite
and that can be described by local strictly positive transformations$^{18}$
is a strong indication that whatever the eventual theory of quantum gravity,
it is efficiently simulatable on a quantum computer.  A particularly 
striking demonstration of this last fact is Freedman's demonstration
that quantum computers can efficiently simulate topological quantum
field theories.$^{42}$

The third interpretation --- that the numbers of bits and ops calculated
here represent the actual memory capacity and number of elementary quantum
logic operations performed by the universe --- is more
controversial.  That the universe registers an amount of information 
equal to the logarithm of its number of accessible states seems reasonable. 
But to say that $2\int E dt/\pi\hbar$ is the number of elementary operations
performed is equivalent to saying that the universe performes an `op' every 
time some piece of it evolves by an average angle (or acquires an
average phase) $\pi/2$ in Hilbert space.
In fact, for a quantum computer, this is a reasonable definition of an
elementary quantum logic operations: to perform 
each quantum logic operation requires energy $E$ to be
made available for a time $\pi\hbar/2E$.   And whenever energy $E$
is available to some degrees of freedom for time $\pi\hbar/2E$, the state of
those degrees of freedom evolves by an angle $\pi/2$ in Hilbert space.
In addition, almost any interaction between degrees of freedom, including
fundamental interactions such as those in the standard model, supports 
quantum logic and quantum computation.$^{12-16, 43}$  Accordingly, it seems 
plausible to identify an elementary quantum logic operation with the local 
evolution of information-carrying degrees of freedom by an angle
of $\pi/2$.  Whether or not this is a reasonable
definition of an elementary operation for the universe as a whole is a
question whose answer will have to await further results on the relationship
between information processing and fundamental physics.

Is the universe a computer?  The answer to this question depends both on the
meaning of `computer' and on the meaning of `is.'  On the one hand,
the universe is certainly {\it not} a digital computer running Linux or 
Windows.  (Or at any rate, {\it not yet}.)  On the other hand, the
universe certainly does represent and process information in a systematic
fashion: as noted above, every spin can be taken to represent a bit,
and every spin flip corresponds to a bit flip.  Almost any interaction
between degrees of freedom suffices to perform universal quantum
logic on those degrees of freedom.$^{12-13}$  Such universal interactions 
include the fundamental interactions of quantum electrodynamics and of
topological quantum field theories.$^{12-16,42-43}$  These results
strongly suggest that the universe as a whole is capable
of universal quantum computation.  In such a `universal' universal
computer, every degree
of freedom in the universe would register information, and
the dynamics of those degrees of freedom would process that information:
the amount of information and the amount of information processing possible
can be calculated using the formulae derived above.    
Even if the universe is not a conventional computer, it can still compute. 

\vfill\eject

\bigskip\noindent{\bf References:}

\item{1.} Landauer, R., {\it Nature}, {\bf 335}, 779-784 (1988).

\item{2.} Lloyd, S., {\it  Nature}, (2000).

\item{3.} Margolus, N., Levitin, L.B., in {\it PhysComp96}, T. Toffoli,
M. Biafore, J. Leao, eds. (NECSI, Boston) 1996; {\it Physica D} {\bf 120}, 
188-195 (1998). 

\item{4.} Peres, A., {\it Quantum Theory:
Concepts and Methods}, (Kluwer, Hingham) 1995.  

\item{5.} Benioff, P., {\it J. Stat. Phys.} {\bf 22}, 563
(1980). 

\item{6.} Feynman, R.P., {\it Int. J. Theor.
Phys.} {\bf 21}, 467 (1982).

\item{7.} Deutsch, D., {\it Proc. Roy. Soc. Lond.} {\bf A 400},
97 (1985). 

\item{8.} Lloyd, S., {\it Science}, {\bf 261}, pp. 1569-1571 (1993).

\item{9.} Shor, P., in 
{\it Proceedings of the 35th Annual Symposium on Foundations
of Computer Science}, S. Goldwasser, Ed., IEEE Computer
Society, Los Alamitos, CA, 1994, pp. 124-134.

\item{10.} Lloyd, S., {\it Sci. Am.} {\bf 273}, 140 (1995).

\item{11.} Divincenzo, D., {\it Science} {\bf 270}, 255 (1995).
 
\item{12.} Lloyd, S., {\it Phys.\ Rev.\ Lett.\/}, {\bf 75},
346-349 (1995).

\item{13.} Deutsch, D., Barenco, A., Ekert, A., {\it Proc.\
Roy.\ Soc.\ A}, {\bf 449}, 669-677 (1995).
 
\item{14.} Cirac, J.I., Zoller, P. {\it Phys. Rev. Lett.} {\bf 74}, 
4091-4094 (1995).

\item{15.} Turchette, Q.A., Hood, C.J., Lange, W., Mabuchi, H.,
Kimble, H.J., {\it Phys. Rev. Lett.}, {\bf 75}, 4710-4713 (1995).

\item{16.} Monroe, C., Meekhof, D.M., King, B.E., Itano, W.M., 
Wineland, D.J., {\it Phys. Rev. Lett.}, {\bf 75}, 4714-4717 (1995).

\item{17.} Grover, L.K., in {\it Proceedings of the 28th Annual ACM Symposium
on the Theory of Computing}, ACM Press, New York, 1996, pp. 212-218.

\item{18.} Lloyd, S.,  {\it Science}, {\bf 273}, 1073-1078 (1996).

\item{19.} Abrams, D., Lloyd, S.,  {\it Phys. Rev. Lett.} {\bf 81},
3992-3995, (1998).
 
\item{20.} Zalka, C., {\it Proc. R. Soc. Lond A. Mat.} {\bf 454}, 
313-322 (1998).

\item{21.} Cory, D.G., Fahmy, A.F., Havel, T.F., 
in {\it PhysComp96, Proceedings of the
Fourth Workshop on Physics and Computation}, T. Toffoli, M. Biafore,
J. Le\~ao, eds., (New England Complex Systems Institute, Boston) 1996.

\item{22.} Gershenfeld, N.A., Chuang, I.L. 
{\it Science} {\bf 275}, pp. 350-356
(1997).

\item{23.} Chuang, I.L., Vandersypen, L.M.K.,
Zhou, X., Leung, D.W., Lloyd, S., {\it Nature} {\bf 393},
143-146 (1998) May, 1998.

\item{24.} Jones, J.A., Mosca, M., Hansen, R.H., {\it Nature}
{\bf 393}, 344-346 (1998). 

\item{25.} Chuang, I.L., Gershenfeld, N., Kubinec, M., {\it Phys. Rev.
Lett.} {\bf 80}, 3408-3411 (1998).

\item{26.} Nakamura, Y., Pashkin, Yu.A., Tsai, J.S., {\it Nature}
{\bf 398}, 305 (1999)

\item{27.} Mooij, J.E., Orlando, T.P., Levitov, L., Tian, L.,
van der Wal, C.H., Lloyd, S., {\it Science} {\bf 285}, 1036-1039
(1999).

\item{28.} Bekenstein, J.D., {\it Phys. Rev. D.} {\bf 23}, 287 (1981).

\item{29.} Bekenstein, J.D., {\it Phys. Rev. Letters\/} {\bf 46}, 623 (1981).

\item{30.} Bekenstein, J.D., {\it Phys. Rev. D.} {\bf 30}, 1669-1679 (1984).

\item{31.} Pagels, H., {\it The Cosmic Code: quantum physics as the
language of nature}, (Simon and Schuster, New York) 1982.

\item{32.} Peebles, P.J.E., {\it Principles of Physical Cosmology},
Princeton University Press, Princeton (1993).

\item{33.} Linde, A., {\it Particle Physics and Inflationary Cosmology},
Harwood Academic, New York (1990).

\item{34.} Liddle, A.R. and Lyth, D.H., {\it Cosmological Inflation
and Large-Scale Structure}, Cambridge University Press, Cambridge
(2000).

\item{35.} Zizzi, P., {\it Entropy} {\bf 2}, 36-39 (2000).  

\item{36.} Zizzi, P., `The Early Universe as a Quantum Growing
Network,' in Proceedings of IQSA Fifth Conference, March 31-April 5,
2001, Cesena-Cesenatico, Italy; Archive gr-qc/0103002.

\item{37.} Ng, J., {\it Phys. Rev. Lett.} {\bf 86}, 2946-2947 (2001).

\item{38.} Eddington, A.S., {\it The Mathematical Theory of
Relativity}, Cambridge University Press, Cambridge (1924).

\item{39.} Dyson, F., {\it Rev. Mod. Phys.} {\bf 51}, 447-460 (1979).

\item{40.} Wheeler, J.A., in {\it Complexity, Entropy, and the Physics
of Information}, W.H. Zurek, ed., Santa Fe Institute Studies in
the Sciences of Complexity volume VIII, Addison Wesley, Redwood City (1988).

\item{41.} Fredkin, E., `Ultimate Limits to Computation,' MIT Laboratory
for Computer Science Seminar (2000).

\item{42.} Freedman, M.H., Kitaev, A., Wang, Z., `Simulation of 
topological quantum field theories by quantum computers,'
quant-ph/0001071.

\item{43.} Freedman, M.H., Larsen, M., Wang, Z., `A modular functor
which is universal for quantum computation,' quant-ph/0001108.

\vfill\eject\end